\documentclass[sigconf,authorversion,nonacm]{acmart}

\usepackage{algorithm}
\usepackage{algpseudocode}
\usepackage{caption}
\usepackage{subcaption}
\usepackage{makecell}
\usepackage{multirow}

\newcommand{\lnand}{\overline{\land}}

\AtBeginDocument{%
  }

\setcopyright{acmcopyright}
\copyrightyear{2018}
\acmYear{2018}
\acmDOI{XXXXXXX.XXXXXXX}

\acmConference[Conference acronym 'XX]{Make sure to enter the correct
  conference title from your rights confirmation emai}{June 03--05,
  2018}{Woodstock, NY}
\acmPrice{15.00}
\acmISBN{978-1-4503-XXXX-X/18/06}

\begin{document}

% \title{Large Circuit Model: An End-to-end Generative Approach to Logic Synthesis}
% \title{Large Circuit Model: End-to-end Logic Synthesis by Predicting the Next Gate}
% \title[Circuit Transformer: End-to-end Logic Synthesis by Predicting the Next Gate]{Circuit Transformer:\\End-to-end Logic Synthesis by Predicting the Next Gate}
\title[Logic Synthesis with Generative Deep Neural Networks]{Logic Synthesis with Generative Deep Neural Networks}
% CTRW: A Generative Neural Rewriter for Logic Synthesis
% Mastering Circuit Optimization with Generative Deep Neural Networks

\author{Xihan Li}
\orcid{0000-0002-7000-7983}
\affiliation{%
  \institution{University College London}
  \city{London}
  \country{UK}
}
\email{xihan.li@cs.ucl.ac.uk}

\author{Xing Li}
\affiliation{%
  \institution{Huawei Noah's Ark Lab}
  \city{Hong Kong}
  \country{China}
}
\email{li.xing2@huawei.com}

\author{Lei Chen}
\affiliation{%
  \institution{Huawei Noah's Ark Lab}
  \city{Hong Kong}
  \country{China}
}
\email{lc.leichen@huawei.com}

\author{Xing Zhang}
\affiliation{%
  \institution{Huawei Noah's Ark Lab}
  \city{Hong Kong}
  \country{China}
}
\email{zhangxing85@huawei.com}

\author{Mingxuan Yuan}
\affiliation{%
  \institution{Huawei Noah's Ark Lab}
  \city{Hong Kong}
  \country{China}
}
\email{Yuan.Mingxuan@huawei.com}

\author{Jun Wang}
\affiliation{%
  \institution{University College London}
  \city{London}
  \country{UK}
}
\email{jun.wang@cs.ucl.ac.uk}


\begin{abstract}
    % While deep learning has dominated multiple domains from board game to conversational agent, its impact on logic circuit design is still limited due to complicated constraints. However, a recent generative deep neural model ``Circuit Transformer'' shed light on this direction, with equivalence-preserving circuit transformation in small scale. In this paper, we further build a rewriting operator based on the model for logic synthesis, named ``\texttt{ctrw}'' (Circuit Transformer Rewriting), including the following proposed techniques: (1) we introduce a two-stage scheme to train Circuit Transformer for logic synthesis, and iteratively improve its optimality via self-imporvement training; (2) we cooperate Circuit Transformer with state-of-the-art rewriting techniques to address the scalability issue, which also allow DAG-aware rewriting to be performed in a guided way. Experimental results on IWLS 2023 contest benchmark demonstrate the effectiveness of our proposed rewriting methods.
    While deep learning has achieved significant success in various domains, its application to logic circuit design has been limited due to complex constraints and strict feasibility requirement. However, a recent generative deep neural model, ``Circuit Transformer'', has shown promise in this area by enabling equivalence-preserving circuit transformation on a small scale. In this paper, we introduce a logic synthesis rewriting operator based on the Circuit Transformer model, named ``\texttt{ctrw}'' (Circuit Transformer Rewriting), which incorporates the following techniques: (1) a two-stage training scheme for the Circuit Transformer tailored for logic synthesis, with iterative improvement of optimality through self-improvement training; (2) integration of the Circuit Transformer with state-of-the-art rewriting techniques to address scalability issues, allowing for guided DAG-aware rewriting. Experimental results on the IWLS 2023 contest benchmark demonstrate the effectiveness of our proposed rewriting methods.
\end{abstract}



%   % \vspace{-10pt}
%   \includegraphics[width=\textwidth,trim={0 8cm 0 4cm},clip]{figures/pipeline.pdf}
%   % \vspace{-10pt}
%   \caption{The logic synthesis pipeline with Circuit Transformer. From left to right, the number of AND ($\wedge$) nodes in the toy AIG network is reduced by one, while the logical functionality remains equivalent.}


\maketitle
% \vspace{-5pt}
\section{Introduction}

\begin{figure*}[t]
    \centering
    \includegraphics[width=1\linewidth,trim={0.7cm 8cm 3.5cm 4cm},clip]{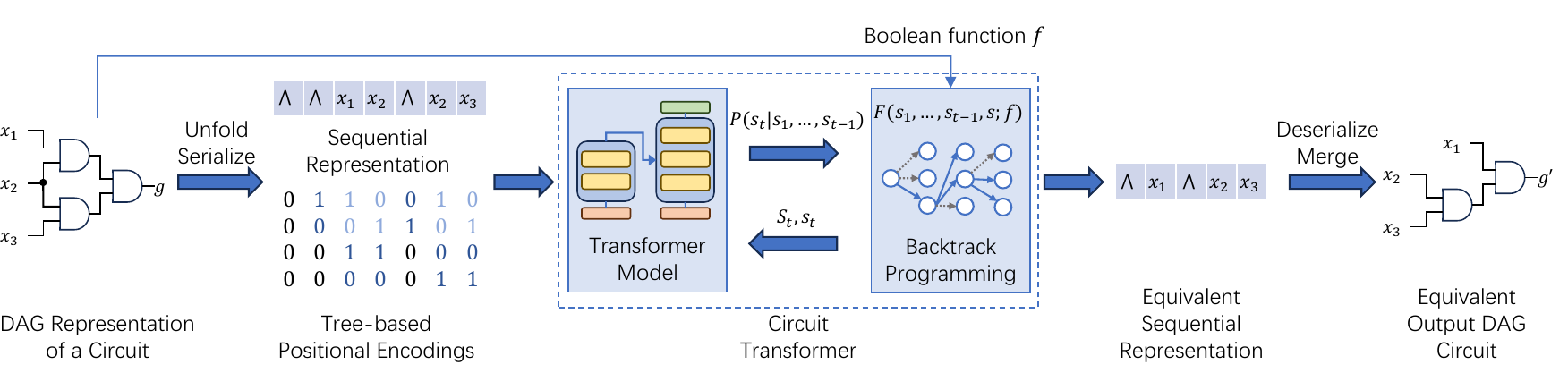}
    \caption{The pipeline of Circuit Transformer that transforms a circuit to a strictly equivalent one.}
    \label{fig:circuit_transformer}
\end{figure*}

% Recent years have witnessed the rapid development of deep learning, with its great success in various fields. Among these fields, two of the most notable ones are the success of deep reinforcement learning on board game, represented by the victory of AlphaGo \cite{alphago} in 2016; and the success of generative neural networks on conversational agent, represented by the release of ChatGPT \cite{chatgpt} in 2022. Given these successes, in fields like board game and conversational agent, people reach consensus that deep-learning based solutions have established a ``dominance'', that is, they are far more advantageous over traditional methods.

In recent years, deep learning has demonstrated rapid advancements and remarkable success across various domains. Two particularly notable achievements include the triumph of deep reinforcement learning in board games, exemplified by AlphaGo's victory in 2016 \cite{alphago}, and the success of generative neural networks in conversational agents, illustrated by the release of ChatGPT in 2022 \cite{chatgpt}. In domains such as board games and conversational agents, there is a consensus that deep learning-based solutions have established a ``dominance'', significantly surpassing traditional methods in terms of performance and capability.

% However, in some other fields like logic circuit design, while there are still lots of deep-learning related works published, the position of deep learning in these fields are not so significant, and traditional methods are still taking the center of the stage. This is usually because of core domain-specific issues that lie in the weak side of deep learning. For logic circuits, one of the core issues is the strict logical constraints that must be satisfied without a single bit of error \cite{ML4EDASurvey2021} (e.g., a synthesized circuit should be precisely equivalent to the original one), while eliminating errors in deep learning is difficult as it is born to be a regression on data. Such a conflict seriously hinders deep learning from realizing its full potential on logic circuit. As a result, most deep learning related works have to avoid direct construction of logic circuits by themselves, but aid traditional methods, which guarantee constraint satisfaction, in constructing better circuits \cite{LSOracle,DRiLLS,BOiLS,Zhu_MLCAD2020,Haaswijk_ISCAS18,chang2023chipgpt,Chip-Chat,liu2023chipnemo}.

However, in fields such as logic circuit design, despite numerous publications on deep-learning related research, deep learning has not yet achieved a dominant position. Traditional methods still play a central role due to domain-specific challenges where deep learning has inherent limitations. For logic circuits, one of the core issues is the need to adhere to strict logical constraints without any margin for error \cite{ML4EDASurvey2021}. For instance, a synthesized circuit must be exactly equivalent to the original, while deep learning, being fundamentally a regression technique, struggles to eliminate errors entirely. This conflict significantly limits the effectiveness of deep learning in logic circuit design. Consequently, most deep learning research in this area focuses on augmenting traditional methods that ensure constraint satisfaction, rather than directly constructing logic circuits. \cite{LSOracle,DRiLLS,BOiLS,Zhu_MLCAD2020,Haaswijk_ISCAS18,chang2023chipgpt,Chip-Chat,liu2023chipnemo}

% In 2024, a new deep neural model ``Circuit Transformer'' \cite{li2024circuit} shed light on tackling the constraint issue of logic circuit. By a specific designed encoding-decoding process, Circuit Transformer allows And-Inverter Graphs (AIGs) to be generated in an analogous way to large language models like ChatGPT, and more importantly, with equivalence constraint precisely preserved. Such a preservation of equivalence is especially appealing to logic synthesis. While up to now it can only process tiny circuits with up to around 30 gates on a customer-grade computer, it do construct circuits from scratch as a fully neural approach, proving that the aforementioned core issue of circuits for deep learning can actually be resolved.

In 2024, the introduction of the Circuit Transformer \cite{li2024circuit} marked a breakthrough in addressing the constraint issues inherent in logic circuit design. By employing a specially designed encoding-decoding process, the Circuit Transformer facilitates the generation of And-Inverter Graphs (AIGs) in a manner analogous to large language models such as ChatGPT. Crucially, this model precisely preserves equivalence constraints. Although the current implementation is limited to processing small circuits comprising up to approximately 30 gates on a consumer-grade computer, it demonstrates the capability to construct feasible circuits from scratch using a fully neural approach. This achievement indicates that the core issue of strict logical constraints in deep learning for circuit design can indeed be overcome, demonstrating a potential paradigm shift in this field.

% Given the core issue resolved, in this paper, we take a further step on Circuit Transformer, showing how this neural model can serve as the cornerstone of a strong operator for logic synthesis. First, to train Circuit Transformer specifically on the task of logic synthesis, we propose a two-stage scheme. The first stage is to train a Circuit Transformer in a supervised way to mimic traditional synthesis tools. The second stage is to improve the optimality of the model with Monte-Carlo Tree Search (MCTS). The general idea is similar to AlphaGo that first trained on human expert moves and then iteratively train on self-play datasets. In such a way, we allow Circuit Transformer to explore and exploit new optimization strategies automatically, without prior experience from human. Moreover, real circuit data is neither not required as only synthetic circuits are used to train the model.

Given that the core issue has been resolved, this paper takes a further step with Circuit Transformer, demonstrating how this neural model can serve as the foundation for rewriting in logic synthesis. To train Circuit Transformer specifically for logic synthesis, we propose a two-stage scheme. In the first stage, the model is trained in a supervised manner to emulate traditional synthesis tools. In the second stage, we iteratively enhance the model's optimality using Monte-Carlo Tree Search (MCTS). This approach is analogous to AlphaGo's training process, which first learned from human expert moves and then improved through iterative training on self-play datasets. This method allows Circuit Transformer to autonomously explore and exploit new optimization strategies without relying on human expertise.

% Then, to address the scalability issue, we cooperate Circuit Transformer with a recent rewrite framework based on fanout-free window \cite{MFFW}, a sub-structure of circuits with one or more outputs. Circuit Transformer is responsible to optimize these fanout-free windows extracted by the rewrite framework, which is usually small in size, rather than optimizing the whole circuit. Meanwhile, from the perspective of sub-graphs in rewriting, compared with enumerating all NPN classes (\cite{mishchenko_dag-aware_2006} with cut size = 4) or exact synthesis (\cite{riener_--fly_2019} with cut size = 6), Circuit Transformer is more scalable as it can process larger sub-graphs (window size = 8 in our case) with multiple outputs, which makes it a good fit as larger considered sub-graphs for rewriting has more potential in creating better optimized networks.

Then, to address the scalability issue, we integrate Circuit Transformer with a recent rewrite framework based on fanout-free windows \cite{MFFW}, which are sub-structures of circuits containing one or more outputs. Instead of optimizing the entire circuit, Circuit Transformer focuses on optimizing these fanout-free windows, which are typically small in size. This approach leverages the strengths of Circuit Transformer to handle these smaller, manageable portions effectively. From a sub-graph rewriting perspective, compared to traditional methods like enumerating all NPN classes with a cut size of 4 \cite{mishchenko_dag-aware_2006} or exact synthesis with a cut size of 6 \cite{riener_--fly_2019}, Circuit Transformer offers greater scalability. It can process larger sub-graphs with a window size of 8 and multiple outputs. This capability makes it particularly well-suited for rewriting, as considering larger sub-graphs holds more potential for creating optimized networks.

% Finally, we show that Circuit Transformer can be adapted to perform DAG-aware rewriting in a natural way. Benefited from its sequential generation process, when we optimize a fanout-free window via Circuit Transformer, we can take the context information (i.e., the nodes which are not in the fanout-free window) into account, and ``guide'' the model to reuse existing nodes during the generation process. More specifically, when a node is generated at a certain step and identified as equivalent to an existing node in the context, the reward at this step will be refined, as if the generated node is replaced by the existing node. In such a way, the cumulative reward of the MDP will correctly reflect the influence of node reuse, and MCTS will be directed to generate circuits that may not be the most compact in size, but leads to more reduced node of the whole circuit after replacement.

Finally, we demonstrate that Circuit Transformer can be adapted for Directed Acyclic Graph (DAG) aware rewriting. Benefiting from a Markov Decision Process (MDP) formulation, Circuit Transformer can incorporate context information (i.e., nodes outside the fanout-free window) during the generation, which allows the model to favor the reuse of existing nodes. Specifically, when a node is generated and identified as equivalent to an existing node in the context, the immediate reward at that step is refined as if the generated node were merged with the existing one. This refinement ensures that the cumulative reward of the MDP accurately reflects the number of added nodes after the merging process. Consequently, Monte-Carlo Tree Search (MCTS) is guided by the reward to generate circuits that may not be the smallest in size but result in a more optimized overall circuit after node merging.

% Together with all the techniques proposed above, we developed a new rewrite operator \texttt{ctrw} (Circuit Transformer Rewriting). At the current stage, while we have only trained a small Circuit Transformer on a customer-grade computer, it still shows impressive performance on IWLS 2023 contest benchmark. Future work includes scaling up the Circuit Transformer model on large-scale AI platform, and improve both efficiency and performance with state-of-the-art AI techniques.

Combining all the techniques proposed above, we developed a new rewrite operator named \texttt{ctrw} (Circuit Transformer Rewriting). Despite currently training a small Circuit Transformer on a consumer-grade computer, \texttt{ctrw} has demonstrated its effectiveness on the IWLS 2023 contest benchmark. Future work will focus on scaling up the Circuit Transformer model using large-scale AI platforms and enhancing both its efficiency and performance with state-of-the-art AI techniques.

\section{Preliminaries}

\subsection{And-Inverter Graph and Fanout-Free Window}

An And-Inverter Graph (AIG) is a data structure to model combinational logic circuits \cite{AIGER}. Let $G = (V, E, PI, PO)$ be an AIG, where $(V, E)$ is a directed acyclic graph, $PI$ is the set of primary input vertices, and $PO$ is the set of primary output vertices. Each vertex $v \in V$ represents an And gate and $v \in PI$ represents a primary input. Edges represent wires and can either be regular or complemented.

Given a set $I$ of $k$ input vertices , a $k$-input fanout-free window (FFW) $G = (V, E, I, O)$ is also an AIG, where $I$ are the input vertices and $O \subset V$ are the output vertices. A fanout-free window should satisfy following fanout-free rules, 
\begin{enumerate}
    \item For each vertex $v \in V$ , $FI(v) \subset V$.
    \item For each vertex $v \in V \ O$, $FO(v) \subset V$.
\end{enumerate}

We refer to \cite{MFFW} for more details.

\subsection{Transformer and Circuit Transformer Model}

The Transformer model \cite{transformer} is a deep learning architecture that revolutionized artificial intelligence, particularly with the emergence of large language models. It typically runs as a generative model that accepts a sequence of tokens (words) as input, and recurrently predict the next token (word) to be generated. Unlike traditional sequential models, transformers process entire sequences simultaneously, allowing for efficient parallelization and capturing long-range dependencies. This architecture, coupled with innovations like multi-head attention and positional encoding, enables transformers to excel in diverse tasks such as natural language processing and image recognition.

The Circuit Transformer model \cite{li2024circuit}, as the name suggests, is a variation of Transformer model for logic circuits. It consists of an encoding scheme that transform an And-Inverter Graph to a trajectory that can be efficiently processed by Transformer models, and a decoding scheme that guarantee the generated circuit to adheres to specified equivalence constraints. It runs typically in a generative way that accepts an encoded AIG as input, and recurrently predict the next node (AND gate or PI), thereby constructing an output AIG step by step. Its pipeline is shown in \autoref{fig:circuit_transformer} and the generation process is shown in \autoref{alg:circuit_transformer}.

\begin{algorithm}[t]
\caption{Processing an AIG with Circuit Transformer}\label{alg:circuit_transformer}
\small
\begin{algorithmic}
\Require An input AIG $G$ to be processed, an optional set of mapping $C$ for the $i$th primary output of the generate AIG, a trained Circuit Transformer $P_{\text{CT}}(g_n | g_{1:n-1}, G_{\text{enc}}, \{TT_i\})$
\Ensure A processed AIG $G'$.
\State Encode $G$ as $G_{\text{enc}}$ following Section 3.1 of \cite{li2024circuit}
\State Initialize $G_{\text{dec}} \leftarrow []$
\While{$t = 0, 1, 2, \dots$}
    \State Compute a probability distribution of nodes
    $$p_t \leftarrow P_{\text{CT}}(g_t | G_{\text{dec}}, G_{\text{enc}}, \{TT_i\})$$
    \State Compute the $t$th node $g_t \leftarrow \arg \max p_t$
    \State \textbf{If} $g_t = \text{EOS}$ \textbf{then} \textbf{break}
    \State Add $g_t$ to $G_{\text{dec}}$
\EndWhile
\State Decode $G_{\text{dec}}$ as $G'$ following Section 3.3 of \cite{li2024circuit}
\State Return $G'$
\end{algorithmic}
\end{algorithm}

\subsection{Markov Decision Process and Monte-Carlo Tree Search}

Markov Decision Processes (MDPs) is a mathematical framework for modeling decision-making problems, in which an agent interacts with an environment over a series of discrete time steps. At each step, the agent chooses an action based on the current state of the environment, leading to a transition to a new state and receiving a reward. By solving MDPs, agents can learn optimal policies (i.e., optimal mappings from state to action) that maximize cumulative rewards over time.

Monte Carlo Tree Search (MCTS) is a heuristic search algorithm renowned for its effectiveness in decision-making under uncertainty. Unlike traditional search methods, MCTS iteratively builds a search tree by sampling potential trajectories and dynamically balancing exploration and exploitation. Through repeated simulations, MCTS efficiently navigates large decision spaces, finding optimal solutions in domains with complex branching factors and uncertain outcomes.

\subsection{Rewriting and DAG-aware Rewriting}

Rewriting is a powerful and widely used area-oriented optimization method that iteratively selects and greedily replaces small-scale sub-graphs with more compact structures. The sub-graphs can be either single-output ($k$-input cuts \cite{mishchenko_dag-aware_2006}) or multi-output ($k$-input fanout-free window \cite{MFFW}).

The idea of DAG-aware rewriting is to replace the aforementioned small-scale subgraphs with ones that can make the whole graph more compact by utilizing existing logic. Such a replacement sub-graph is not necessary to be smaller than the original one. Existing approaches compute multiple candidate replacements offline \cite{mishchenko_dag-aware_2006} or on-the-fly \cite{riener_--fly_2019}, then enumerate each candidate during the rewriting process, and select the one with best gain to replace the original sub-graph. Information about existing logic are not utilized when the candidate replacements are generated.

\section{Methods}

\subsection{Train a Circuit Transformer for Logic Synthesis}\label{sec:train_a_circuit_transformer}

\begin{algorithm}[t]
\caption{Random generation of a $k$-input, $l$-output AIG}\label{alg:random_circuit}
\small
\begin{algorithmic}
\Require Number of input $k$, number of output $l$, number of steps $M_{\text{step}}$.
\Ensure A randomly generated AIG with $k$ inputs and $l$ outputs.
\State $S \leftarrow \{I_0, I_1, \dots, I_{k-1}\}$
\For{$i = 1, 2, \dots, M_{\text{step}}$}
    \State Create an AND node $a_i$
    \State Randomly sample two nodes $c_0, c_1 \in S$ without replacement
    \State Set the first input of $a_i$ as $c_0$ or $\overline{c_0}$ randomly
    \State Set the second input of $a_i$ as $c_1$ or $\overline{c_1}$ randomly
    \State Add $a_i$ to $S$
\EndFor
\State Return an AIG with $I_0, I_1, \dots, I_{k-1}$ as primary inputs and $a_{M-l+1}, a_{M-l+2}, \dots, a_{M}$ as primary outputs.
\end{algorithmic}
\end{algorithm}

% In this section, we train a Circuit Transformer model to perform logic synthesis on small-scale AIGs in a fully generative way. The input of the model is an encoded AIG to be optimized, and the neural model generate the synthesized AIG, which is equivalent to the original one and expected to be more compact in size, step by step by recurrently predicting the next node of the AIG.

In this section, we train a Circuit Transformer model to perform logic synthesis on small-scale AIGs in a fully generative manner. The input to the model is an encoded AIG that needs optimization. The neural model then generates the synthesized AIG step by step by recurrently predicting the next node of the AIG. The output AIG is equivalent to the original but expected to be more compact in size.

% To train the Circuit Transformer model for logic synthesis, we propose a two-stage training scheme. The first stage is to train the model in a supervised way on pairs of data (original circuit, optimized circuit), in which the original circuits are randomly generated, and the optimized circuits are generated by traditional logic synthesis libraries such as \texttt{resyn2} in ABC \cite{abc}. In such a way, we initialize the Circuit Transformer model by mimicking the strategy of traditional synthesis tools. To generate a $k$-input, $l$-output AIG in a random way, we initial a set of nodes $S = \{I_0, I_1, \dots, I_{k-1}\}$ in which $I_0, I_1, \dots, I_{k-1}$ are the input nodes. Then, we iteratively create an AND node $a_i$, randomly choose two nodes $c_0, c_1 \in S$ as the two inputs of $a_i$ (the two wires connecting them are randomly set to be regular or complemented), and add $a_i$ to $S$. After iterating for $M$ steps, we select the last $l$ nodes as the primary outputs of the generated AIG. The detailed process is shown in \autoref{alg:random_circuit}. Canonicalization techniques \cite{MFFW} can be applied to drop AIGs with existing caononicalizations in the dataset. This is to ensure that each randomly generated AIG in the dataset has a unique structure, thereby enhancing the diversity of the dataset. 

To train the Circuit Transformer model for logic synthesis, we propose a two-stage training scheme. In the first stage, the model undergoes supervised training using pairs of data in the form <original circuit, optimized circuit>. The original circuits are randomly generated, while the optimized circuits are produced using conventional logic synthesis libraries, such as \texttt{resyn2} in ABC \cite{abc}. This approach initializes the Circuit Transformer by emulating the strategies used by traditional synthesis tools. 

To generate a $k$-input, $l$-output AIG in a random manner, we are inspired by the Boolean chain \cite{knuth2020art}, starting by initializing a set of nodes \( S = \{ I_0, I_1, \dots, I_{k-1} \} \), where \( I_0, I_1, \dots, I_{k-1} \) are input nodes. We then iteratively create AND nodes \( a_i \). For each \( a_i \), we randomly select two nodes \( c_0, c_1 \in S \) as its inputs, setting the connecting wires to be either regular or inverted at random. The AND node \( a_i \) is subsequently added to \( S \). After \( M \) iterations, the last $l$ nodes are designated as the primary outputs of the generated AIG. The detailed process is outlined in \autoref{alg:random_circuit}.

Canonicalization techniques \cite{MFFW} are applied to filter out AIGs with pre-existing canonicalizations in the dataset. This ensures that each randomly generated AIG in the dataset has a unique structure, thereby enhancing dataset diversity.

% It is worth noting that, while realistic data is usually believed to be always better than synthetic one for training, we found it hard to train a workable model using cuts or fanout-free windows extracted from realistic circuits as training data. We hypothesize that the reason is that the gaps between original and optimized circuits are mostly narrow in real cuts or fanout-free windows, making the model hard to learn effective node-reducing strategies.

It is worth noting that, although realistic data is generally considered superior to synthetic data for training, we found it challenging to train an effective model using cuts or fanout-free windows extracted from realistic circuits. We hypothesize that this difficulty arises because the differences between original and optimized circuits are typically minimal in real cuts or fanout-free windows, making it hard for the model to learn effective node-reduction strategies.

\subsection{Iterative Self-Improvement Training}

\begin{figure*}[t]
    \centering
    \includegraphics[width=0.7\linewidth,trim={2cm 4cm 9cm 3.5cm},clip]{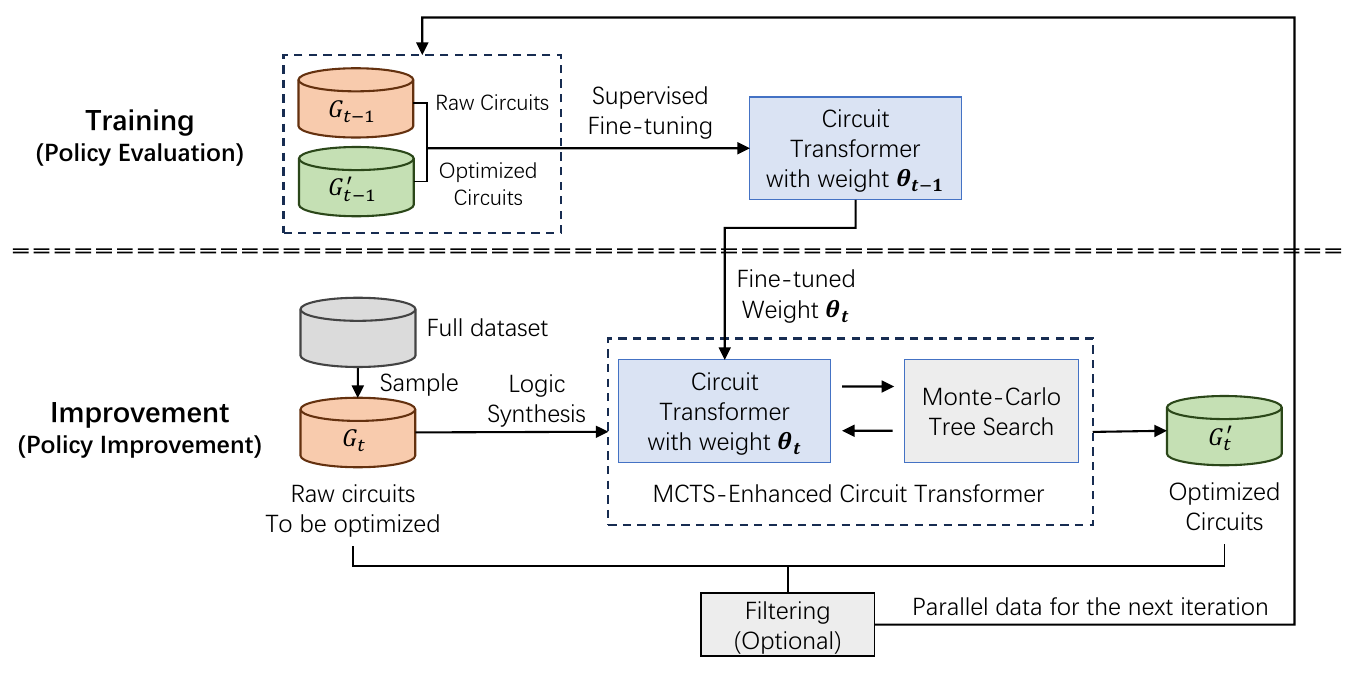}
    \caption{The process of Iterative Self-Improvement Training. Two stages are included in a single iteration: (1) Training: the Circuit Transformer is supervisedly fine-tuned by circuit pairs $<G_{t-1}, G'_{t-1}>$, with its weight updated from $\theta_{t-1}$ to $\theta_t$; (2) Improvement: a batch of raw circuits $G_t$ is sampled from the full dataset, and processed by MCTS-enhanced Circuit Transformer with weight $\theta_t$ to generate further optimized circuits $G'_t$.}
    \label{fig:iterative_training}
\end{figure*}

\begin{figure*}[t]
    \centering
    \begin{subfigure}[b]{0.4\linewidth}
         \centering
         \includegraphics[width=0.9\textwidth,trim={0.5cm 3cm 20cm 5cm},clip]{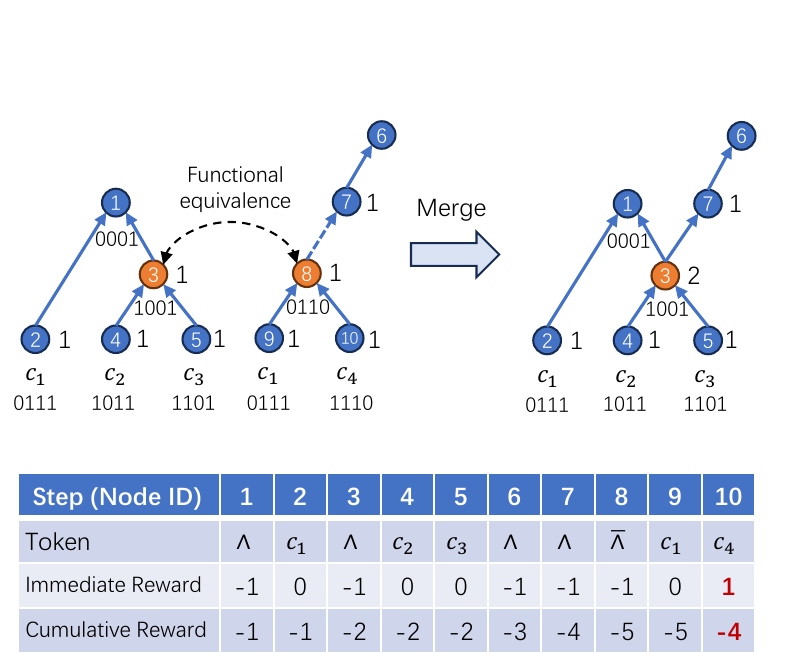}
         \caption{In step 10, the functionality of node 8 is fully determined ($c_1 \land c_4$), which is functionally equivalent to node 3 ($c_2 \land c_3$). Therefore, the two nodes are merged and the number of AND nodes is reduced by one, which is added to the reward of step 10. The cumulative reward up to step 10 is -4, which correctly reflects the number of added AND nodes after merging (node 1, 3, 6, 7).}
         \label{fig:local_merging}
    \end{subfigure}
    \hspace{5pt}
    \begin{subfigure}[b]{0.55\linewidth}
         \centering
         \includegraphics[width=0.8\textwidth,trim={17cm 3cm 1.5cm 5cm},clip]{figures/merge.pdf}
         \caption{In step 13, the functionally of node 11 (truth table: 1100) and node 7 (truth table: 1000) are determined, and node 7 is identified to be functionally equivalent to the green node in the whole circuit. Therefore node 7 is dereferenced as it is replaced by the green node, reducing the number of AND nodes by two (node 7 and 11), which is added to the reward of step 13. The cumulative reward up to step 13 is -3, which correctly reflects the number of added AND nodes after the DAG-aware merging (node 1, 3, 6).}
         \label{fig:contextual_merging}
    \end{subfigure}
    \caption{Examples showing how the reward is refined in the sequential generation process, so that the cumulative reward correctly reflects the total number of added AND nodes.}
    \label{fig:merge}
\end{figure*}

% Given an initial Circuit Transformer, the second stage is to iteratively improve the optimality of the model for generating more compact circuits. To achieve this, we attach an immediate reward function to each step in the stepwise circuit generation process discribed in \autoref{alg:circuit_transformer}, so as to regard it as a Markov Decision Process, in which the state, action and reward in step $t$ are defined as follows:
Given an initial Circuit Transformer, the second stage focuses on iteratively enhancing the model's ability to generate more compact circuits. To achieve this, we attach an immediate reward function to each step of the circuit generation process described in \autoref{alg:circuit_transformer}. This transforms the stepwise process into a Markov Decision Process (MDP), where the state, action, and reward at step $t$ are defined as follows:
\begin{itemize}
    % \item State: $G_{\text{dec}}, G_{\text{enc}}, \{TT_i\}$. (note that only $G_{\text{dec}}$ will change over steps)
    \item State: the generated token sequence $g_1, \dots, g_t$
    \item Action: all the valid choices of $g_{t+1}$.
    % \item Reward: -1 if $g_t$ is an AND gate, or 0 if $g_t$ is a primary input. If there are one or more gates that is merged with previous ones due to functional equivalence, then the reward should also plus the number of merged gates.
    \item Immediate Reward:
        \begin{equation}
        R(g_1, \dots, g_t, g) = \Delta + \begin{cases}
            -1, &g = \land \text{ or } g = \lnand \\
            0, &\text{otherwise}
        \end{cases}
    \end{equation}
    in which $\Delta$ reflects the refinement due to equivalent node merging. An example can be found in \autoref{fig:local_merging}.
\end{itemize}

In this way, the cumulative reward is the negative of number of AND nodes in the generated AIG, and the Circuit Transformer model serves as a policy, mapping from state $g_1, \dots, g_t$ to action $g_{t+1}$. Consequently, the task of logic synthesis, with its core objective of minimizing the number of AND nodes, can be reformulated as solving this MDP.

% To solve this MDP, we need to find an approach to iteratively improve the policy, which is served by the Circuit Transformer here. While a neurally represented policy of an MDP can usually be improved by deep reinforcement learning (DRL) algorithms such as Q-learning or policy gradient, the Transformer-based architecture hinders us to do so in a trivial way. We observe that a direct application of online DRL algorithms on this MDP results in non-convergence, and no significant improvement with offline DRL algorithms. This may due to the incredibly sensitivity of DRL algorithms over network architectures, and significant non-stationary during training, which is discussed in \cite{li2023a}. 
To solve this MDP, we need an effective method to iteratively improve the policy, as represented by the Circuit Transformer. While policies in MDPs can often be enhanced using deep reinforcement learning (DRL) algorithms such as Q-learning or policy gradient methods, the unique features of the Transformer-based architecture prevent straightforward application of these algorithms. Our observations indicate that directly applying online DRL algorithms to this MDP results in non-convergence, and offline DRL methods do not yield significant improvements either. This behavior may be attributed to the extreme sensitivity of DRL algorithms to network architectures and the substantial non-stationarity during training, as discussed in \cite{li2023a}. 

% To stabilize the training process, we propose to leverage Monte-Carlo Tree Search to improve the policy. When deciding the $t$-th node $g_t$, instead of simply selecting the node with the highest probability from Circuit Transformer, we build a search tree to evaluate multiple choices of $g_t$ (i.e., actions in MDP) via simulation, and Circuit Transformer model serves as a guidance to decide the priority of exploration. The detailed process is shown in \autoref{alg:mcts}. Then, we propose Iterative Self-improvement Training, which is to iteratively fine-tune the Circuit Transformer model on the improved data pairs. In each iteration, we first fine-tune the Circuit Transformer on data pairs (original circuits, optimized circuits) in a supervised way. Then we randomly sample a set of circuits from the full dataset, and process them via MCTS-enhanced Circuit Transformer to generate enhanced optimized circuits. The sampled circuits and the enhanced optimized circuits then serve as the data pairs in the next iteration for fine-tuning. The detailed process is shown in \autoref{alg:iterative_training}.

To stabilize the training process, first, we propose leveraging Monte-Carlo Tree Search (MCTS) to improve the policy. When deciding on the \( t \)-th node \( g_t \), instead of merely selecting the node with the highest probability from the Circuit Transformer, we build a search tree to evaluate multiple choices of \( g_t \) (i.e., actions in the MDP) through simulation. The Circuit Transformer model acts as a guide to prioritize exploration. The detailed process is outlined in \autoref{alg:mcts}. 

Next, we introduce Iterative Self-improvement Training to iteratively fine-tune the Circuit Transformer model. In each iteration, the Circuit Transformer is first fine-tuned on data pairs <original circuits, optimized circuits> in a supervised manner. Then, we randomly sample a set of circuits from the complete dataset and process them through the MCTS-enhanced Circuit Transformer to generate further optimized circuits. These sampled circuits and their corresponding enhanced optimized circuits then serve as the data pairs for fine-tuning in the next iteration. The detailed iterative training process is depicted in \autoref{fig:iterative_training} and \autoref{alg:iterative_training}.

\begin{algorithm}[t]
\caption{Logic synthesis with MCTS-enhanced Circuit Transformer}\label{alg:mcts}
\small
\begin{algorithmic}
\Require An input AIG $G$ to be processed, an optional truth table $TT_i$ for the $i$th primary output of the generate AIG, a trained Circuit Transformer $P_{\text{CT}}(g_n | g_{1:n-1}, G_{\text{enc}}, \{TT_i\})$, number of MCTS steps $M_{\text{step}}$, number of MCTS playouts per step $M_{\text{playout}}$
\Ensure A processed AIG $G'$.
\Procedure{PUCT}{an MCTS node $x$}
    \For{$a$ in $x$'s all child nodes}
        \State $s_a \leftarrow \frac{a.\text{total\_value}}{a.\text{visited}} + a.\text{prob}\sqrt{\frac{x.\text{visited}}{1 + a.\text{visited}}}$
    \EndFor
    \State \textbf{Return} $\arg \max_a s_a$
\EndProcedure
\Procedure{MCTS}{$P_{\text{CT}}$}  \Comment{See \cite{alphago_zero} for details}
    \State Create a root MCTS node $r$
    \For{$i = 1, 2, \dots, M_{\text{playout}}$}
        \State (Selection) Starting from $r$, iteratively selects a child node via PUCT algorithm until reaching a leaf node $l$.
        \State (Expansion) Evaluate $l$ via $P_{\text{CT}}$ (i.e., create all child nodes $a$ for $l$, and assign $a$.prob for each child via $P_{\text{CT}}$), and select a child node via PUCT.
        \State (Simulation) Run the generation process in \autoref{alg:circuit_transformer} via $P_{\text{CT}}$ until reaching EOS or the maximum \#(iter), and get the cumulative reward $v$
        \State (Backpropagation) Update the ``visited'' and ``total\_value'' attributes of the MCTS nodes from $p$ to $r$ in a backward pass with $v$.
    \EndFor
    \State \textbf{Return} An action of $r$ in which the maximum cumulative reward appeared in the corresponding branch.
\EndProcedure
\State Encode $G$ as $G_{\text{enc}}$ following Section 3.1 of \cite{li2024circuit}
\State Initialize $G_{\text{dec}} = []$
\While{$t = 0, 1, 2, \dots$}
    \If{$t < M_{\text{step}}$}
        \State Compute the $t$-th node $g_t \leftarrow \text{MCTS}(P_{\text{CT}})$
    \Else
        \State $g_t \leftarrow \arg \max P_{\text{CT}}(g_t | G_{\text{dec}}, G_{\text{enc}}, \{TT_i\})$
    \EndIf
    \State \textbf{If} $g_t = \text{EOS}$ \textbf{then} \textbf{break}
    \State Add $g_t$ to $G_{\text{dec}}$
\EndWhile
\State Decode $G_{\text{dec}}$ as $G'$ following Section 3.3 of \cite{li2024circuit}
\State \textbf{Return} $G'$
\end{algorithmic}
\end{algorithm}

\begin{algorithm}[t]
\caption{Iterative improvement training process}\label{alg:iterative_training}
\small
\begin{algorithmic}
\Require The Circuit Transformer model $P_{\text{CT}}$ to be fine-tuned, number of iterations $M_{\text{iter}}$, dataset size per iteration $M_{\text{size}}$, a pre-generated dataset $D$, number of MCTS playouts $M_{\text{playout}}$
\Ensure An enhanced Circuit Transformer model.
\State $P_{\text{CT}}^0 \leftarrow P_{\text{CT}}$
\For{$i = 0, 1, \dots, M_{\text{iter}} - 1$}    
    \State $D_i \leftarrow []$
    \While{$|D_i| < M_{\text{size}}$}
        \State Sample $(G, G') \in D$
        \State Do a random NPNP transformation to $G$.
        \State Run \autoref{alg:circuit_transformer} with $P_{\text{CT}}^i$ on $G$ to get $G''_1$
        \State Run \autoref{alg:mcts} with $P_{\text{CT}}^i$ on $G$ to get enhanced AIG $G''_2$
        \If{$|G''_2| < |G''_1|$}
            \State Add $(G, G''_2)$ to $D_i$
        \EndIf
    \EndWhile
    \State Fine-tune $P_{\text{CT}}^i$ on dataset $D_i$ to get $P_{\text{CT}}^{i+1}$
\EndFor
\State \textbf{Return} $P_{\text{CT}}^{M_{\text{iter}} - 1}$.
\end{algorithmic}
\end{algorithm}

\subsection{Cooperate Circuit Transformer with Fanout-Free Window Rewriting}\label{sec:ctrw}

While the Circuit Transformer model we trained in the previous section can already perform end-to-end logic synthesis with decent performance, it is of limited scalability. In exchange for the decoding efficiency of Transformer models, Circuit Transformer's encoding scheme allows ``unfolding'' with significant redundancy, which hinders it to process very large circuits (as the encoded sequence will be prohibitively long). On a customer-grade computer equipped with a single high-end GPU, we are able to train a Circuit Transformer that can process AIGs of up to around 30 AND nodes. Even if resorting to large-scale AI training clusters (which we are actually working on), we would not expect Circuit Transformer to process AIGs with thousands of nodes in a single shot. 

Therefore, to address the scalability issue, we also resort to so-called peephole optimization, which partitions a circuit into small sub-circuits that can be replaced by a more compact implementation. More specifically, we leverage the fanout-free window (FFW) rewriting framework proposed in \cite{MFFW}, and apply our trained Circuit Transformer with MCTS enhancement to replace each fanout-free window by a more compact one. We name such a rewrite operator as ``\textit{Circuit Transformer Rewriting}'' (\texttt{ctrw}). A filter will be applied to candidate FFWs, to ensure that the circuits are within the trained model's capacity after encoding.

\begin{figure}[t]
  \centering
  \begin{subfigure}[b]{0.43\linewidth}
     \centering
     \includegraphics[width=\textwidth,trim={1.5cm 12.5cm 22cm 1cm},clip]{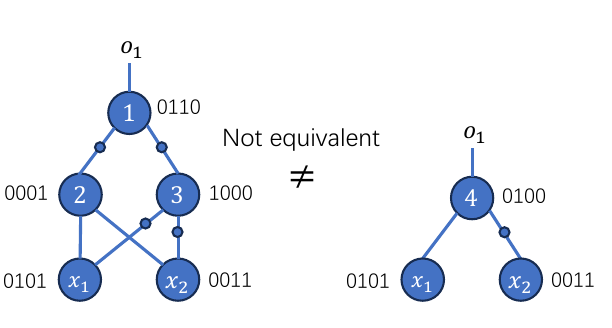}
     \caption{Two AIGs which are not equivalent.}
     \label{fig:xor}
  \end{subfigure}
  \hspace{3pt}
  \begin{subfigure}[b]{0.54\linewidth}
     \centering
     \includegraphics[width=\textwidth,trim={1cm 6cm 21cm 6.5cm},clip]{figures/equivalence.pdf}
     \caption{The first AIG, which serves as a sub-graph in this case, can be replaced by the second one.}
     \label{fig:xor_in_context}
  \end{subfigure}
  \caption{An example showing how a sub-circuit can be replaced by a non-equivalent one, without changing the functionality of the whole circuit.}
  \label{fig:equivalence}
\end{figure}

Moreover, Circuit Transformer enables additional flexibility for the rewriting process. For the partition-and-replacement scheme of rewriting frameworks, an important observation is that, while the new implementation should not change the original functionality of the whole circuit, the new implementation itself is not necessary to be logical equivalent to the sub-circuit it replaces \cite{reichl_circuit_2023}. An example is shown in \autoref{fig:equivalence}. Traditional methods like offline replacement computation \cite{mishchenko_dag-aware_2006} or on-the-fly exact synthesis \cite{riener_--fly_2019} have to preserve such logical equivalence between sub-circuits, resulting in a loss of flexibility. A recent study \cite{reichl_circuit_2023} allows such kind of flexibility via a specific encoding of Quantified Boolean Formulas, and Circuit Transformer provides a more straightforward approach with its capability of preserving arbitrary equivalence constraints.

To illustrate, we first introduce how a Circuit Transformer preserves the equivalence between input and output circuits in the previous section. For a circuit with $k$ inputs and $l$ outputs, this is done by specifying $2^k \times l$ constraints so as to fix the truth table of the outputs. For example, for the first AIG in \autoref{fig:xor} consisting of three AND nodes, when feeding it into a Circuit Transformer to generate a new equivalent AIG, four constraints will be specified: $(x_0 = 0, x_1 = 0 \rightarrow o_1 = 0)$, $(x_0 = 1, x_1 = 0 \rightarrow o_1 = 1)$, $(x_0 = 0, x_1 = 1 \rightarrow o_1 = 1)$, $(x_0 = 1, x_1 = 1 \rightarrow o_1 = 0)$.

Then, when we apply a Circuit Transformer to rewrite a $k$-input sub-graph within a larger AIG of $K$ inputs, we specify the constraints in a more ``globalized'' way. That is, we resort to the truth table of the input and output nodes, with respect to the larger AIG (so the size of the truth tables are $2^K$ instead of $2^k$). For example, for the first sub-graph in \autoref{fig:xor_in_context} within a 5-nodes AIG, we utilize the truth table of two outputs (node 4 and 5) and one inputs (node 1) for the whole graph, and specify two constraints for the Circuit Transformer: $(x_0 = 0, x_1 = 0 \rightarrow o_1 = 0)$ and $(x_0 = 1, x_1 = 1 \rightarrow o_1 = 0)$. In such a way, we introduce additional flexibility by reducing the number of equivalence constraints from four to two, leaving the other two input cases ($x_0 = 1, x_1 = 0$ and $x_0 = 0, x_1 = 1$) as don't cares. Once a Circuit Transformer generates an AIG satisfying these two constraints, it can replace the original sub-graph without changing the whole circuit, as the truth table of the output nodes within the whole graph are preserved.

It also worth attention that cycles may appear when replacing a multi-output sub-circuit by a new one, which is also observed in \cite{reichl_circuit_2023}. For our method, we simply check whether a cycle appears after each replacement, and revert the change if a cycle is detected.

\subsection{Circuit Transformer for Guided DAG-aware Rewriting}\label{sec:DAG-aware}

The idea of DAG-aware rewriting bases on another important observation that, while the aim of rewriting is to make the whole circuit more compact with less nodes, the new implementation itself is not necessary to be more compact than the sub-circuit it replaces. This is because the new implementation may heavily reuse already existing logic in the whole circuit, which will not be counted after the replacement. Previous works \cite{mishchenko_dag-aware_2006,riener_--fly_2019} implement DAG-aware rewriting via generating multiple candidate circuits for a single replacement, and select the one that leads to a minimal number of nodes of the whole circuit after replacement. Such methods are less ``guided'' as the structure of the whole circuit is not considered during the generation of candidate circuits. 

Circuit Transformer, however, is able to consider such context information during the MCTS enhancement described in \autoref{alg:mcts}, so as to perform a ``guided search'' that favor the reuse of existing logic, and minimize the number of nodes of the whole circuit. This is done by refining the definition of reward in the MDP proposed in \autoref{sec:train_a_circuit_transformer}, to correctly reflect the node reduction of the whole circuit after replacement, by taking existing nodes' reuse into consideration. More specifically, we first de-reference the sub-graph to be replaced, then during the sequential generation process in \autoref{alg:mcts}, when we found that the functionality of a generated AND node can be fully determined (i.e., no nodes it depends on are still waiting to be generated), we will check whether the node is functionally equivalent to an existing node in the whole circuit. If such a node exists, we will do a ``mock de-reference'' of the generated node to see how many nodes can be reduced by replacing this generated node with the existing one. Then, we add the number of ``mock de-referenced'' nodes to the reward in the current step. In such a way, the cumulative reward of the MDP will correctly reflect the influence of node reuse, and MCTS will be directed to generate circuits that may not be the most compact in size, but leads to more reduced node of the whole circuit after replacement. An example is shown in \autoref{fig:contextual_merging}. We argue such a ``DAG-aware generation process'' can be more efficient than traditional techniques in finding an optimal circuit structure for replacement.

\section{Experiments}

\begin{table*}[]
    \centering
    \footnotesize
    \resizebox{\textwidth}{!}{%
\setlength{\tabcolsep}{3pt}
\begin{tabular}{ll|lrrlrr|lrrlrrlrrlrr}
\hline
\multicolumn{2}{c|}{Benchmark} & \multicolumn{6}{c|}{Existing Rewriting Methods} & \multicolumn{12}{c}{Proposed Rewriting Method} \\ \hline
\multicolumn{1}{l|}{\multirow{2}{*}{Name}} & \multicolumn{1}{c|}{\multirow{2}{*}{Size}} & \multicolumn{3}{c|}{\begin{tabular}[c]{@{}c@{}}Drw Rewriting\\ (abc)\end{tabular}} & \multicolumn{3}{c|}{\begin{tabular}[c]{@{}c@{}}MFFW Rewriting\\ (python)\end{tabular}} & \multicolumn{3}{c|}{\begin{tabular}[c]{@{}c@{}}ctrw w/o Iterative Self-\\ Improvement Training\end{tabular}} & \multicolumn{3}{c|}{ctrw} & \multicolumn{3}{c|}{ctrw with MCTS} & \multicolumn{3}{c}{\begin{tabular}[c]{@{}c@{}}ctrw with MCTS \& guided\\ DAG-aware rewriting\end{tabular}} \\ \cline{3-20} 
\multicolumn{1}{l|}{} & \multicolumn{1}{c|}{} & Size & Improv. & \multicolumn{1}{r|}{Time} & Size & Improv. & Time & Size & Improv. & \multicolumn{1}{r|}{Time} & Size & Improv. & \multicolumn{1}{r|}{Time} & Size & Improv. & \multicolumn{1}{r|}{Time} & Size & Improv. & Time \\ \hline
\multicolumn{1}{l|}{ex12} & 12 & 12 & 0.00\% & \multicolumn{1}{r|}{0.01} & 12 & 0.00\% & 5.25 & 12 & 0.00\% & \multicolumn{1}{r|}{13.55} & 12 & 0.00\% & \multicolumn{1}{r|}{16.07} & 10 & 16.67\% & \multicolumn{1}{r|}{294.28} & 10 & 16.67\% & 242.95 \\
\multicolumn{1}{l|}{ex17} & 17 & 17 & 0.00\% & \multicolumn{1}{r|}{0.01} & 17 & 0.00\% & 3.51 & 17 & 0.00\% & \multicolumn{1}{r|}{0.80} & 17 & 0.00\% & \multicolumn{1}{r|}{0.92} & 17 & 0.00\% & \multicolumn{1}{r|}{356.27} & 16 & 5.88\% & 417.22 \\
\multicolumn{1}{l|}{ex32} & 33 & 29 & 12.12\% & \multicolumn{1}{r|}{0.01} & 30 & 9.09\% & 2.63 & 31 & 6.06\% & \multicolumn{1}{r|}{14.65} & 27 & 18.18\% & \multicolumn{1}{r|}{16.10} & 27 & 18.18\% & \multicolumn{1}{r|}{2618.13} & 24 & 27.27\% & 2748.27 \\
\multicolumn{1}{l|}{ex23} & 26 & 24 & 7.69\% & \multicolumn{1}{r|}{0.01} & 24 & 7.69\% & 49.33 & 24 & 7.69\% & \multicolumn{1}{r|}{27.68} & 24 & 7.69\% & \multicolumn{1}{r|}{42.62} & 24 & 7.69\% & \multicolumn{1}{r|}{4246.21} & 22 & 15.38\% & 2534.06 \\
\multicolumn{1}{l|}{ex89} & 14 & 13 & 7.14\% & \multicolumn{1}{r|}{0.01} & 13 & 7.14\% & 4.63 & 13 & 7.14\% & \multicolumn{1}{r|}{3.21} & 13 & 7.14\% & \multicolumn{1}{r|}{3.00} & 12 & 14.29\% & \multicolumn{1}{r|}{686.61} & 12 & 14.29\% & 674.20 \\
\multicolumn{1}{l|}{ex99} & 43 & 41 & 4.65\% & \multicolumn{1}{r|}{0.01} & 38 & 11.63\% & 3.53 & 40 & 6.98\% & \multicolumn{1}{r|}{39.83} & 35 & 18.60\% & \multicolumn{1}{r|}{34.15} & 34 & 20.93\% & \multicolumn{1}{r|}{3208.47} & 35 & 18.60\% & 3809.46 \\
\multicolumn{1}{l|}{ex30} & 28 & 24 & 14.29\% & \multicolumn{1}{r|}{0.01} & 24 & 14.29\% & 7.17 & 25 & 10.71\% & \multicolumn{1}{r|}{12.41} & 22 & 21.43\% & \multicolumn{1}{r|}{16.65} & 22 & 21.43\% & \multicolumn{1}{r|}{2050.01} & 21 & 25.00\% & 1942.07 \\
\multicolumn{1}{l|}{ex07} & 39 & 27 & 30.77\% & \multicolumn{1}{r|}{0.01} & 27 & 30.77\% & 1.63 & 27 & 30.77\% & \multicolumn{1}{r|}{30.18} & 27 & 30.77\% & \multicolumn{1}{r|}{7.81} & 27 & 30.77\% & \multicolumn{1}{r|}{712.62} & 27 & 30.77\% & 691.69 \\
\multicolumn{1}{l|}{ex96} & 37 & 19 & 48.65\% & \multicolumn{1}{r|}{0.01} & 19 & 48.65\% & 3.58 & 19 & 48.65\% & \multicolumn{1}{r|}{9.93} & 19 & 48.65\% & \multicolumn{1}{r|}{4.31} & 19 & 48.65\% & \multicolumn{1}{r|}{806.41} & 19 & 48.65\% & 659.56 \\
\multicolumn{1}{l|}{ex41} & 39 & 25 & 35.90\% & \multicolumn{1}{r|}{0.01} & 20 & 48.72\% & 0.91 & 24 & 38.46\% & \multicolumn{1}{r|}{6.51} & 21 & 46.15\% & \multicolumn{1}{r|}{24.68} & 21 & 46.15\% & \multicolumn{1}{r|}{2566.13} & 20 & 48.72\% & 2679.71 \\
\multicolumn{1}{l|}{ex57} & 44 & 40 & 9.09\% & \multicolumn{1}{r|}{0.01} & 40 & 9.09\% & 158.45 & 40 & 9.09\% & \multicolumn{1}{r|}{112.32} & 40 & 9.09\% & \multicolumn{1}{r|}{107.29} & 38 & 13.64\% & \multicolumn{1}{r|}{16863.31} & 38 & 13.64\% & 11793.32 \\
\multicolumn{1}{l|}{ex21} & 48 & 39 & 18.75\% & \multicolumn{1}{r|}{0.01} & 33 & 31.25\% & 27.75 & 38 & 20.83\% & \multicolumn{1}{r|}{21.39} & 35 & 27.08\% & \multicolumn{1}{r|}{37.97} & 35 & 27.08\% & \multicolumn{1}{r|}{4062.32} & 33 & 31.25\% & 3692.64 \\
\multicolumn{1}{l|}{ex53} & 71 & 63 & 11.27\% & \multicolumn{1}{r|}{0.01} & 53 & 25.35\% & 123.37 & 53 & 25.35\% & \multicolumn{1}{r|}{208.82} & 54 & 23.94\% & \multicolumn{1}{r|}{235.94} & 53 & 25.35\% & \multicolumn{1}{r|}{21801.94} & 51 & 28.17\% & 19789.97 \\
\multicolumn{1}{l|}{ex35} & 75 & 57 & 24.00\% & \multicolumn{1}{r|}{0.01} & 53 & 29.33\% & 11.36 & 54 & 28.00\% & \multicolumn{1}{r|}{169.56} & 41 & 45.33\% & \multicolumn{1}{r|}{115.18} & 44 & 41.33\% & \multicolumn{1}{r|}{9810.85} & 34 & 54.67\% & 7157.42 \\
\multicolumn{1}{l|}{ex14} & 53 & 47 & 11.32\% & \multicolumn{1}{r|}{0.01} & 42 & 20.75\% & 1.31 & 48 & 9.43\% & \multicolumn{1}{r|}{11.65} & 42 & 20.75\% & \multicolumn{1}{r|}{24.61} & 39 & 26.42\% & \multicolumn{1}{r|}{2566.86} & 38 & 28.30\% & 2577.84 \\
\multicolumn{1}{l|}{ex49} & 66 & 61 & 7.58\% & \multicolumn{1}{r|}{0.01} & 60 & 9.09\% & 304.63 & 60 & 9.09\% & \multicolumn{1}{r|}{249.72} & 60 & 9.09\% & \multicolumn{1}{r|}{285.02} & 61 & 7.58\% & \multicolumn{1}{r|}{31421.68} & 56 & 15.15\% & 35465.11 \\
\multicolumn{1}{l|}{ex79} & 59 & 46 & 22.03\% & \multicolumn{1}{r|}{0.01} & 30 & 49.15\% & 0.98 & 31 & 47.46\% & \multicolumn{1}{r|}{67.46} & 30 & 49.15\% & \multicolumn{1}{r|}{86.21} & 28 & 52.54\% & \multicolumn{1}{r|}{6356.74} & 28 & 52.54\% & 5996.43 \\
\multicolumn{1}{l|}{ex44} & 81 & 70 & 13.58\% & \multicolumn{1}{r|}{0.01} & 68 & 16.05\% & 2.38 & 67 & 17.28\% & \multicolumn{1}{r|}{120.17} & 65 & 19.75\% & \multicolumn{1}{r|}{169.98} & 56 & 30.86\% & \multicolumn{1}{r|}{9831.79} & 52 & 35.80\% & 12344.27 \\
\multicolumn{1}{l|}{ex95} & 70 & 53 & 24.29\% & \multicolumn{1}{r|}{0.01} & 42 & 40.00\% & 3.13 & 43 & 38.57\% & \multicolumn{1}{r|}{203.44} & 43 & 38.57\% & \multicolumn{1}{r|}{236.67} & 43 & 38.57\% & \multicolumn{1}{r|}{22409.89} & 40 & 42.86\% & 19073.88 \\
\multicolumn{1}{l|}{ex06} & 73 & 69 & 5.48\% & \multicolumn{1}{r|}{0.01} & 62 & 15.07\% & 65.12 & 61 & 16.44\% & \multicolumn{1}{r|}{44.97} & 60 & 17.81\% & \multicolumn{1}{r|}{57.12} & 58 & 20.55\% & \multicolumn{1}{r|}{8963.9} & 55 & 24.66\% & 7096.53 \\
\multicolumn{1}{l|}{ex90} & 95 & 82 & 13.68\% & \multicolumn{1}{r|}{0.01} & 72 & 24.21\% & 8.71 & 80 & 15.79\% & \multicolumn{1}{r|}{171.38} & 68 & 28.42\% & \multicolumn{1}{r|}{142.22} & 64 & 32.63\% & \multicolumn{1}{r|}{14827.17} & 48 & 49.47\% & 9050.18 \\
\multicolumn{1}{l|}{ex04} & 77 & 64 & 16.88\% & \multicolumn{1}{r|}{0.01} & 63 & 18.18\% & 18.95 & 66 & 14.29\% & \multicolumn{1}{r|}{101.95} & 59 & 23.38\% & \multicolumn{1}{r|}{93.34} & 53 & 31.17\% & \multicolumn{1}{r|}{8131.76} & 49 & 36.36\% & 6717.23 \\ \hline
\multicolumn{2}{l|}{Avg. Improv.} &  & 15.42\% & \multicolumn{1}{l|}{} &  & 21.16\% & \multicolumn{1}{l|}{} &  & 18.55\% & \multicolumn{1}{l|}{} &  & 23.23\% & \multicolumn{1}{l|}{} &  & 26.02\% & \multicolumn{1}{r|}{} &  & 30.19\% &  \\ \hline
\end{tabular}%
}
    \caption{Results on 22 small cases of IWLS 2023 contest dataset. Time is measured in seconds.}
    \label{tab:IWLS2023}
\end{table*}

To employ the mature development ecosystem of deep learning, We implement our method fully in Python with TensorFlow. While the inference speed of Circuit Transformer limits the scalability of our implementation on customer-grade computers, it can be significantly scaled and accelerated on modern large-scale AI clusters with thousands of GPUs, which is our ongoing work. Before that, we trained and evaluated a relatively small Circuit Transformer on a customer-grade computer with dual NVIDIA RTX 4090 GPUs, and release preliminary results in this section to show the effectiveness of our proposed rewriting method.

To train the Circuit Transformer model, we build a dataset containing 12 million randomly generated 8-input, 2-output AIGs. The supervised signal (i.e., the synthesized circuits) are generated by the \texttt{resyn2} command in ABC \cite{abc}. We restrict that the length of the encoded sequence for each AIG should be no more than 200, all the 8 inputs should appear in the AIG, and each AIG has an unique canonicalization. 

The details of the Circuit Transformer model are as follows. We employed an encoder-decoder architecture following \cite{transformer}, each with 12 attention layers. The embedding width and the size of feedforward layer are set as 512 and 2048, leading to 88,225,800 total parameters. The implementation is based on \cite{tensorflowmodelgarden2020}. The vocabulary size is 20 (8 POs and AND gate with their inverse, plus [EOS] and [PAD]). Batch size and learning rate are set as 96 and $10^{-3}$ respectively.  The maximum length of the input and output sequence are set as 200 and 100. For the first supervised stage, We trained the model for 1 million batches. For the second iterative enhancement stage, we first set $M_{\text{size}} = 10^6, M_{\text{step}} = 3, M_{\text{playout}} = 10$ for 5 iterations without the filtering condition in line 9 of \autoref{alg:iterative_training}, and then set $M_{\text{size}} = 10^5, M_{\text{step}} = 1, M_{\text{playout}} = 10$ for 24 iterations with the filtering condition. Note that all these parameters are selected to be particularly small so as to fit the limited computational capability of the computer we used. 

To evaluate the optimization capability of our newly proposed Circuit Transformer Rewrite (\texttt{ctrw}) operator, we conduct experiments on 22 cases in IWLS 2023 contest benchmark that are within the computational capability of the computer we used to run \texttt{ctrw} in a reasonable time. The truth tables in the benchmarks are processed by the ABC command sequence \texttt{read\_truth -xf
ex00.truth; collapse; sop; strash; dc2;} suggested in \cite{mishchenko_iwls_2022} to generate the initial AIG circuits to be rewritten. 

The following rewriting techniques are included as baselines:
\begin{itemize}
    \item \textbf{Drw rewriting} \cite{mishchenko_dag-aware_2006}: the latest and best version of cut rewriting in ABC, which is of 4-inputs for each cut.
    \item \textbf{MFFW rewriting} \cite{MFFW}: a recent maximum fanout-free window rewriting with $k=8$. MFFWs are optimized via \texttt{resyn2} in ABC. Note that it is implemented in Python as the base of \texttt{ctrw}, which impacts its running efficiency.
\end{itemize}

For our proposed Circuit Transformer Rewriting, to evaluate the effectiveness of the iterative improvement training stage in \autoref{alg:iterative_training}, we report the results on two models, the initial model supervisedly trained on random circuits, and the enhanced model with iterative self-improvement training. To evaluate the additional flexibility of Circuit Transformer for rewriting that are proposed in \autoref{sec:ctrw} and \autoref{sec:DAG-aware}, we report the result with and without these additional flexibility. For all the \texttt{ctrw} cases with MCTS, we set $M_{\text{step}} = 10$ and $M_{\text{playout}} = 10$. 

% The results are shown in \autoref{tab:IWLS2023}. All the AIGs generated by ctrw have passed equivalence checking with \texttt{cec} in ABC, empirically proving that the proposed neural generative method is capable of generating strictly feasible circuits. We observe that ctrw with enhanced model is significantly better than the one with initial model and also outperforms the MFFW baseline, demonstrating the effectiveness of the proposed iterative self-improvement training. While with significant increase of time cost, MCTS also improves over direct generation, and ctrw with guided DAG-aware rewriting significantly improves over the one without it, demonstrating its effectiveness.
The results are presented in \autoref{tab:IWLS2023}. All of the AIGs generated by \texttt{ctrw} have successfully passed equivalence checking using \texttt{cec} in ABC, empirically demonstrating that the proposed neural generative method is capable of producing strictly feasible circuits. 

We observe that \texttt{ctrw} with the enhanced model significantly outperforms the version with the initial model and also surpasses the MFFW baseline, which highlights the effectiveness of the proposed iterative self-improvement training. Despite the increased time cost, MCTS also shows improvement over direct generation. Furthermore, \texttt{ctrw} with guided DAG-aware rewriting markedly improves over the version without it, further showcasing its effectiveness.

%all \texttt{ctrw}-based results are significantly better than \texttt{mffw}, which means that even a vanilla Circuit Transformer can be more optimal than \texttt{resyn2} on optimizing fanout free windows, once it is enhanced by MCTS. Results of \texttt{ctrw2} are significantly better than \texttt{ctrw1}, indicating that our iterative improvement training in \autoref{sec:train_a_circuit_transformer} is effective. \texttt{ctrw} results with \texttt{-c} option are significantly better than those without \texttt{-c}, indicating that additional flexibility proposed in \autoref{sec:ctrw} and \autoref{sec:DAG-aware} is also effective. While not so significant, \texttt{ctrw} results with \texttt{-c -d} option are worse than those with \texttt{-c} option, indicating that the database technique in \cite{MFFW} can be detrimental on the additional flexibility provided by \texttt{-c}.

\section{Conclusion}

% In this work, we proposed the first logic synthesis operator based on generative deep neural networks. It not only successfully generated strictly feasible circuits, but also effective in reducing the size of the circuit. The application of Circuit Transformer model enables additional flexibility in rewriting, which shows the beauty of cooperation between traditional optimization techniques and modern deep learning models. This paper opens up the direction of generative-AI based logic synthesis, and our recent focus is on scaling up the Circuit Transformer based on large-scale AI platform, and improve both the efficiency and effectiveness of the model with state-of-the-art AI techniques.
In this work, we introduced the first logic synthesis operator powered by generative deep neural networks. This operator not only successfully generated strictly feasible circuits but also demonstrated significant effectiveness in reducing circuit size. The application of the Circuit Transformer model provides additional flexibility in rewriting, showcasing the synergy between traditional optimization techniques and modern deep learning models.

This paper paves the way for generative AI-based approaches in logic synthesis. Moving forward, our focus will be on scaling up the Circuit Transformer using large-scale AI platforms and enhancing both the efficiency and effectiveness of the model with state-of-the-art AI techniques.

\bibliographystyle{ACM-Reference-Format}
\bibliography{references}

\end{document}